# The ϵ-capacity of a gain matrix and tolerable disturbances: Discrete-time perturbed linear systems

O.Zakary, M.Rachik
*University Hassan II-Casablanca-Faculty of sciences Ben M'sik -Department of Mathematics, BP. 7955 Sidi Othmane, Casablanca, Morocco*

***Abstract:*** *Discrete-time linear systems with perturbed initial state are considered. A disturbance that infects the initial state is said to be ϵ-tolerable if the corresponding output signal is relatively insensitive to their effects. In this paper, we will define a new set that characterize each gain matrix K and the associated feedback control law $u_i = Kx_i$, this set will be called the ϵ-capacity of the gain matrix K. The set of all possible gain matrix that makes the system insensitive to all disturbances is noted Φ. The characterization of Φ is investigated, and we propose an algorithmic approach that allows to determine if a control law is belongs to Φ or not. Numerical simulations are given.*

## I. Introduction

Generally, in practice of control systems, we have to shed light on the existence of disturbances, due to the principle of action and counter-action which exists between a system and its environment. There has a lot of work on this topic [1-3], it is the case of systems with perturbed parameters [4]. For instance, we can cite the works of Dahleh et al., in Reference [5] and Wang et al., in Reference [6], where the authors studied respectively the rejection of persistent bounded disturbances, and the exponential filtering for uncertain Markovian jump time-delay systems with nonlinear disturbances. In the mathematical modeling, disturbances lead to the emergence of undesired parameters. Moreover, these parameters could decrease the performance of the system, thus in the worst case destabilize it. In this paper, we are concerned with disturbances that infect the initial state for discrete-time linear system, such disturbances are said to be ϵ-tolerable if the associated output is insensitive to their effects, i.e. the effects of these disturbances are relatively tolerable. In [11], an application of $H_\infty$ control attenuating initial-state uncertainties to the magnetic bearing where the authors examine the $H_\infty$ DIA Disturbance and Initial state uncertainty Attenuation) control problem for a class of linear time-invariant systems. For a given scalar $\epsilon > 0$, we will define the ϵ-capacity of a gain matrix K a set of all initial state that satisfies a certain inequalities with respect to ϵ. We will show that for a given $\epsilon > 0$, if the ϵ-capacity of some gain matrix K verifies some conditions then the system will be insensitive to the effects of all disturbances when the control law $u_i = Kx_i$ is used, i.e. all disturbances will be ϵ-tolerable. In this paper, the ϵ-capacity of some gain matrix K, that makes the system insensitive to the effects of all disturbances, will be called an admissible ϵ-capacity. We are interested with the investigation of the set of all gain matrix those have an admissible ϵ-capacity. More precisely, we consider the following example. A linear, controlled, discrete-time system described by

$$\begin{cases} x_{i+1} = Ax_i + Bu_i \\ x_0 = \alpha\tau_0 + \beta \in \mathbb{R}^n \end{cases} \quad (1)$$

where $\beta = \sum_{j=1}^n \beta_j e_j \in \mathbb{R}^n$ and $\alpha \in \mathbb{R}$ are unknown and unavoidable disturbances that infect the initial state $\tau_0 \in \mathbb{R}^n$, $(e_i)_i, i = 1, \ldots, n$, is the canonical basis of $\mathbb{R}^n$. The associated output function is:

$$y_i = Cx_i \in \mathbb{R}^p$$

where A, B and C are real matrices of appropriate dimension.
The control law is

$$u_i = Kx_i \in \mathbb{R}^m$$

We will propose a technique to determine among these controls law which makes the system insensitive the effects of these disturbances. A disturbance $(\alpha, \beta) \in \mathbb{R}^n \times \mathbb{R}$ is said to be ϵ-tolerable if:

$$\left\|\frac{\partial y_i}{\partial \alpha}\right\| \leq \epsilon \text{ and } \left\|\frac{\partial y_i}{\partial \beta_j}\right\| \leq \epsilon \ \forall j = 1, \ldots, n; \ \forall i \geq 0 \quad (2)$$

Some of the control objectives are to stabilize the system and to maintain its output trajectory within the domain of constraints. The constraints may be summarized by a single set inclusion

$$(y_i) \in Y(\alpha, \beta, \epsilon) = \left\{ (y_i) \in \mathbb{R}^p / \left\|\frac{\partial y_i}{\partial \alpha}\right\| \leq \epsilon \text{ and } \left\|\frac{\partial y_i}{\partial \beta_j}\right\| \leq \epsilon \ \forall j = 1, \ldots, n; \ \forall i \geq 0 \right\} \quad (3)$$

If these constraints are violated for any $i \geq 0$, serious damage may happen. We say that a gain matrix K have an admissible $\epsilon$-capacity $\Theta_\epsilon(K)$, if the output of the system never exceed the specified constraints (3). With (1) and (3), it is desired to determine the set $\Phi$ of all gain matrices K those have an admissible $\epsilon$-capacity, to be explicit:

$$\Phi = \left\{ K \in \mathbb{R}^{m \times n} / \left\|\frac{\partial y_i}{\partial \alpha}\right\| \leq \epsilon \text{ and } \left\|\frac{\partial y_i}{\partial \beta_j}\right\| \leq \epsilon \ \forall j = 1, \ldots, n; \ \forall i \geq 0 \right\}$$

The outline of the rest of the manuscript is: Section 2, the goal of this paper is stated and the required assumptions are discussed, and the main theoretical results are developed in Section 3, while the algorithmic determination for the $\epsilon$-capacity $\Theta_\epsilon(K)$ for each gain matrix $K$ will presented in Section 4, and the simulation results and the concluding remarks are given in Section 5.

## II. Problem Statement

The linear, controlled, discrete-time systems considered in this paper have the following form

$$\begin{cases} x_{i+1} = Ax_i + Bu_i \\ x_0 \end{cases}$$

where $x_0 = \alpha \tau_0 + \beta \in \mathbb{R}^n$, $\beta = \sum_{j=1}^n \beta_j e_j \in \mathbb{R}^n$ and $\alpha \in \mathbb{R}$ are unknown and unavoidable disturbances that infect the initial state $\tau_0 \in \mathbb{R}^n$, $(e_i)_i, i = 1, \ldots, n$, is the canonical basis of $\mathbb{R}^n$.
The associated output function is:

$$y_i = Cx_i \in \mathbb{R}^p$$

where A, B and C are real matrices of appropriate dimension.

**Definition 1**. A disturbance $(\alpha, \beta) \in \mathbb{R} \times \mathbb{R}^n$ is said to be $\epsilon$-tolerable if the corresponding output satisfies the following condition:

$$\left\|\frac{\partial y_i}{\partial \alpha}\right\| \leq \epsilon \text{ and } \left\|\frac{\partial y_i}{\partial \beta_j}\right\| \leq \epsilon \ \forall j = 1, \ldots, n; \ \forall i \geq 0$$

Otherwise $(\alpha, \beta)$ is said $\epsilon$-intolerable.

Which means that the output function is insensitive to the effects of the $\epsilon$-tolerable disturbances, but in the case of $\epsilon$-intolerable disturbances, the desired performance can be lost.
The control law

$$u_i = Kx_i \in \mathbb{R}^m$$

is introduced in order to reduce the effects of these $\epsilon$-intolerable disturbances and/or makes the system insensitive to the effects of all disturbances that infect the initial state. To characterize the utility of each control law, we have the following definition

**Definition 2**. For a given $\epsilon > 0$, and a gain matrix $K \in \mathbb{R}^{m \times n}$, the set

$$\Theta_\epsilon(K) = \{x \in \mathbb{R}^n / \|C(A + BK)^i x\| \leq \epsilon \ \forall i \geq 0\}$$

will be called the $\epsilon$-capacity of the gain matrix $K$.

One can note that $\Theta_\epsilon(K)$, if we consider $(\alpha, \beta) = (1,0) \in \mathbb{R} \times \mathbb{R}^n$ and the set of the constraints $Y(\epsilon) = \{(y_i) \in \mathbb{R}^p / \|y_i\| \leq \epsilon \ \forall i \geq 0\}$, is the set of all initial state such that the corresponding output function satisfies the constraints $(y_i) \in Y(\epsilon)$, in the literature, $\Theta_\epsilon(K)$ is called the maximal output admissible set [7]. The concept of admissible set is very important in the field of analysis and control of linear and non linear systems.

Many papers worked on the construction of admissible sets [7-10]. We can cite the work of Kenji Hirata and Yoshito Ohta in [9], where the authors present necessary and sufficient conditions for fulfilling the specified state and control constraints for a class of nonlinear systems, and they propose an explicit procedure to determine the maximal output admissible set. In [10], an efficient algorithm is provided to compute the maximal set of admissible initial states in the case of autonomous linear discrete-time systems subject to linear constraints.

In this paper, we are interested in studying some applications of this concept to linear discrete-time systems with perturbed initial state. And we will show that if this set have some properties, then the system will be insensitive to the effects of all disturbances under the corresponding control law. One can note that the $\epsilon$-capacity of some gain matrix is described by an infinite number of inequalities, as in [7], the maximal output admissible set is finitely determined, i.e. is described by a finite number of inequalities under some assumptions, in this paper, we show that the controllability of (A,B) and the observability of (A+BK,C) are sufficient for reducing this writing to a finite number of inequalities, which will allow us to use an algorithmic procedure for computing $\Theta_\epsilon(K)$.

**Definition 3**. If a control law $u_i = Kx_i$ makes the system insensitive to the effects of all disturbances, we say that K have an admissible $\epsilon$-capacity $\Theta_\epsilon(K)$.

Our goal is to characterize the set $\Phi$ of all gain matrices $K$ those have an admissible $\epsilon$-capacity $\Theta_\epsilon(K)$, to be explicit:

$$\Phi = \{K \in \mathbb{R}^{m \times n} / (y_i) \in Y(\alpha, \beta, \epsilon), y_i \text{ is the corresponding output}\}$$

Where

$$Y(\alpha, \beta, \epsilon) = \left\{(y_i) \in \mathbb{R}^p / \left\|\frac{\partial y_i}{\partial \alpha}\right\| \leq \epsilon \text{ and } \left\|\frac{\partial y_i}{\partial \beta_j}\right\| \leq \epsilon \ \forall j = 1, \ldots, n; \ \forall i \geq 0\right\}$$

is the set of all output functions which are relatively insensitive to the effects of all disturbances.

In the next section, we will develop the main results that will allow us to assimilate the connection between the set $\Theta_\epsilon(K)$ and the set $\Phi$, and some necessary mathematical preliminaries will be given.

## III. Preliminary Results

First, we will describe the system considered, a linear, controlled, discrete-time system

$$\begin{cases} x_{i+1} = Ax_i + Bu_i \\ x_0 = \alpha\tau_0 + \beta \in \mathbb{R}^n \end{cases} \quad (4)$$

and the associated output function is :
$$y_i = Cx_i \in \mathbb{R}^p \quad (5)$$

the control law is :
$$u_i = Kx_i \in \mathbb{R}^m$$

where the state variable $x_i \in \mathbb{R}^n$ and A, B, C and K are respectively (n, n), (n, m), (p, n) and (m, n) matrices, and $\tau_0 \in \mathbb{R}^n$ is given, , $\beta = \sum_{j=1}^n \beta_j e_j \in \mathbb{R}^n$ and $\alpha \in \mathbb{R}$ are unknown disturbances that infect the initial state, knowing that they are supposed inevitable, $(e_i)_i, i = 1, \ldots, n$, is the canonical basis of $\mathbb{R}^n$. We will say that a disturbance $(\alpha, \beta)$ is tolerable if, for every $i \geq 0$ the corresponding output function $y_i$ is not influenced by the disturbance's effects. That could be mathematically translated by the variation of $y_i$ is independent of $(\alpha, \beta)$ i.e. the partial derivate of $y_i$ with respect to $\alpha$ and with respect to $\beta$ almost vanishes. Given a positive parameter $\epsilon$, a disturbance $(\alpha, \beta)$ is said to be $\epsilon$-tolerable if:

$$\left\|\frac{\partial y_i}{\partial \alpha}\right\| \leq \epsilon \text{ and } \left\|\frac{\partial y_i}{\partial \beta_j}\right\| \leq \epsilon \ \forall j = 1, \ldots, n; \ \forall i \geq 0 \quad (6)$$

In the case of the $\epsilon$-intolerable disturbances, it is desired to find a control law such that the corresponding output function satisfies (6). Then the principal goal is to determine the set $\Phi$ of all gain matrices that makes the system insensitive to the effects of all disturbances $(\alpha, \beta)$, to be explicit:

$$\Phi = \left\{K \in \mathbb{R}^{m \times n} / \left\|\frac{\partial y_i}{\partial \alpha}\right\| \leq \epsilon \text{ and } \left\|\frac{\partial y_i}{\partial \beta_j}\right\| \leq \epsilon \ \forall j = 1, \ldots, n; \ \forall i \geq 0\right\} \quad (7)$$

However, the set $\Phi$ is defined by an infinite number of inequalities, which may be impossible to obtain. We proof that under hypothesis, this difficulty can be overcome and the set $\Phi$ will be described by a finite number of inequalities which allows us to use an algorithmic procedure to characterize it.
Equation (5) can be written as follows:

$$y_i = C(A + BK)^i(\alpha \tau_0 + \beta)$$

In fact

$$x_{i+1} = Ax_i + Bu_i$$
$$x_{i+1} = (A + BK)x_i$$
$$x_i = C(A + BK)^i x_0$$
$$x_i = (A + BK)^i(\alpha \tau_0 + \beta)$$

then

$$y_i = C(A + BK)^i(\alpha \tau_0 + \beta)$$

And

$$\begin{cases} \frac{\partial y_i}{\partial \alpha} = C(A + BK)^i \tau_0 \\ \frac{\partial y_i}{\partial \beta_j} = C(A + BK)^i e_j \end{cases} \quad (8)$$

Using (7) and (8), we can write $\Phi$ as follows:

$$\Phi = \{K \in \mathbb{R}^{m \times n} / \|C(A + BK)^i \tau_0\| \leq \epsilon \text{ and } \|C(A + BK)^i e_j\| \leq \epsilon \ \forall j = 1, \ldots, n; \ \forall i \geq 0\} \quad (9)$$

In the following remark, we will show the interest of introducing the set $\Theta_\epsilon(K)$ in the characterization of $\Phi$. For that, consider $S = \{1, \ldots, n\}$ a set of indices.

**Remark 4**. It is clear that

$$\Phi = \{K \in \mathbb{R}^{m \times n} / \tau_0 \in \Theta_\epsilon(K) \text{ and } e_j \in \Theta_\epsilon(K) \ \forall j \in S\} \quad (10)$$

In fact, let $y_i$ be the output function corresponding to an initial state $x_0 = \alpha \tau_0 + \beta$, the gain matrix K have an admissible $\epsilon$-capacity that makes the disturbance $(\alpha, \beta)$ $\epsilon$-tolerable if:

$$(y_i) \in Y(\alpha, \beta, \epsilon) = \left\{(y_i) \in \mathbb{R}^p / \left\|\frac{\partial y_i}{\partial \alpha}\right\| \leq \epsilon \text{ and } \left\|\frac{\partial y_i}{\partial \beta_j}\right\| \leq \epsilon \ \forall j = 1, \ldots, n; \ \forall i \geq 0\right\}$$

then it is clear from (9) and (10) that:

$$\left\|\frac{\partial y_i}{\partial \alpha}\right\| \leq \epsilon; \forall i \geq 0 \Leftrightarrow \|C(A + BK)^i \tau_0\| \leq \epsilon; \forall i \geq 0 \Leftrightarrow \tau_0 \in \Theta_\epsilon(K)$$

and

$$\left\|\frac{\partial y_i}{\partial \beta_j}\right\| \leq \epsilon \ \forall j = 1, \ldots, n; \ \forall i \geq 0 \Leftrightarrow \|C(A + BK)^i e_j\| \leq \epsilon, \forall j \in S; \forall i \geq 0 \Leftrightarrow e_j \in \Theta_\epsilon(K), \forall j \in S$$

Therefore the system (4) is insensitive to the disturbances $\alpha$ (respectively $\beta$) if and only if $\tau_0 \in \Theta_\epsilon(K)$ (respectively $e_j \in \Theta_\epsilon(K) \ \forall j \in S$).
We note that the $\epsilon$-capacity of K is also defined by an infinite number of inequalities. We will establish sufficient conditions which allow us to describe it by a finite number of inequalities, so $\Phi$.
In order to characterize the set $\Theta_\epsilon(K)$, we introduce, without loss of generality, the following notations:

$$\Theta = \Theta_\epsilon(K) = \{x \in \mathbb{R}^n / \|C\widetilde{A}^i x\| \leq \epsilon \ \forall i \geq 0\}$$

And

$$\Theta_k = \{x \in \mathbb{R}^n / \|C\tilde{A}^i x\| \leq \epsilon, \forall i = 0, \ldots, k\}$$

where

$$\tilde{A} = A + BK$$

**Remark 5**. For every integer $k \geq 0$ we have

$$\Theta \subset \Theta_{k+1} \subset \Theta_k$$

**Proposition 6**. *The $\epsilon$-capacity $\Theta$ of some gain matrix $K$ is:*
i) *Convex*
ii) *Symmetric*
iii) *Closed*
iv) *Contains the origin in its interior*

**Proof**: It is obvious from the definition of $\Theta$.

**Definition 7**. The set $\Theta$ is said to be finitely determined, if there exists an integer $k$ such that:
$$\Theta = \Theta_k$$

The stop condition of the proposed algorithm in the next section is based on the following proposition.

**Proposition 8**. *The set $\Theta$ is finitely determined if and only if $\Theta_{k+1} = \Theta_k$ for some $k \geq 0$.*

**Proof**: $\Theta$ is finitely determined, so there exist an integer $k \geq 0$ such that
$$\Theta = \Theta_k$$
the remark 5 above implies that
$$\Theta = \Theta_{k+1} = \Theta_k$$

Conversely, $\Theta_{k+1} = \Theta_k$ for some $k \geq 0$, let $x \in \Theta_k$ then

$$\|C\tilde{A}^{k+1} x\| \leq \epsilon$$

we deduce that
$$\tilde{A}x \in \Theta_k, \forall x \in \Theta_k$$
and by iteration we have
$$\tilde{A}^i x \in \Theta_k, \forall i \geq 0$$
then
$$x \in \Theta$$
that implies
$$\Theta_k \subset \Theta$$
and we know that $\Theta \subset \Theta_k$ for every $k \geq 0$, which complete the proof.

**Proposition 9**. *If $\|\tilde{A}\| \leq \alpha$ for some $\alpha \in ]0,1[$ then $\Theta$ is finitely determined.*

**Proof**: Let $\epsilon > 0$
$$\|\tilde{A}\| \leq \alpha \Rightarrow \|C\tilde{A}^i x\| \leq \gamma_i \|x\|, \forall i \geq 0$$
where $\gamma_i = \alpha^i \|C\|$.
Since
$$\gamma_i \xrightarrow{i \to \infty} 0$$
then there exists an integer $k_0$ such that

$$\|C\tilde{A}^i x\| \leq \epsilon, \forall i \geq k_0; \forall x \in \mathbb{R}^n$$

then

$$\|C\tilde{A}^{k_0+1}x\| \leq \epsilon; \forall x \in \Theta_{k_0}$$

And by remark 5 and proposition 8 we have that
$$\Theta = \Theta_{k_0+1} = \Theta_{k_0}$$
and $\Theta$ is finitely determined.

**Remark 10**. One can note that the asymptotic stability of $\tilde{A}$ is a special case of the Proposition 9.

**Theorem 11**. *If $\tilde{A}$ is asymptotically stable ( $|\lambda| < 1$ for every $\lambda$ eigenvalue of $\tilde{A}$), then there exists an integer $k_0$ such that the output function $y_i$ is not sensitive to disturbance $\beta$ for every $i > k_0$.*

**Proof**: The asymptotic stability of $\tilde{A}$ implies that there exists an integer $k_0$ such that
$$\|C\tilde{A}^{k_0}\| \leq \epsilon$$
then
$$\|C\tilde{A}^i e_j\| \leq \epsilon, \forall j \in S; \forall i \geq k_0$$

Using (9), we have
$$\left\|\frac{\partial y_i}{\partial \beta_j}\right\| \leq \epsilon, \forall j \in S; \forall i \geq k_0$$

**Remark 12**. If the pair $(A, B)$ is controllable, then we can choose K such that the matrix $\tilde{A}$ is asymptotically stable. So the controllability is a sufficient condition for the theorem 11.

**Theorem 13**. *Suppose the following assumptions hold:*
*i) the pair $(\tilde{A}, C)$ is observable*
*ii) the pair $(A, B)$ is controllable*
*Then $\Theta$ is finitely determined.*

**Proof**: By the observability of $(\tilde{A}, C)$, the rank of the matrix $\Lambda$ is n, where
$$\Lambda = \begin{bmatrix} C \\ C\tilde{A} \\ C\tilde{A}^2 \\ \vdots \\ C\tilde{A}^{n-1} \end{bmatrix}$$
which implies that $\Lambda^T \Lambda$ is invertible, so there exists a $\rho \geq 0$ such that
$$\rho \|x\|^2 \leq \langle \Lambda^T \Lambda x, x \rangle, \forall x \in \mathbb{R}^n \qquad (11)$$
and we have
$$\Lambda x \in \overbrace{B(0,\epsilon) \times ... \times B(0,\epsilon)}^{n \text{ times}}, \forall x \in \Theta_{n-1}$$

where $B(0,\epsilon) = \{x \in \mathbb{R}^n / \|x\| \leq \epsilon\}$.
Since $\overbrace{B(0,\epsilon) \times ... \times B(0,\epsilon)}^{n \text{ times}}$ is bounded, then there exists a $\lambda \geq 0$ such that
$$\langle \Lambda^T \Lambda x, x \rangle \leq \lambda \|x\|, \forall x \in \Theta_{n-1} \qquad (12)$$

from (11) and (12), there exists $\gamma \geq 0$ such that
$$\|x\| \leq \gamma, \forall x \in \Theta_{n-1} \qquad (13)$$
the pair $(A, B)$ is controllable, so there exists K such that the matrix $\tilde{A} = A + BK$ is asymptotically stable, then there exists an integer $k_0 \geq n - 1$ such that
$$\|C\tilde{A}^{k_0}\| \leq \frac{\epsilon}{\gamma}$$

from (13) we have
$$\|C\tilde{A}^{k_0}x\| \leq \epsilon, \forall x \in \Theta_{n-1}$$
which implies that $\Theta_{n-1} \subset \Theta_{k_0}$ and we know that $\Theta_{k_0} \subset \Theta_{n-1}$ because $k_0 \geq n - 1$ then $\Theta_{k_0} = \Theta_{n-1}$ and $\Theta$ is finitely determined.

*The ϵ-capacity of a gain matrix and tolerable disturbances: Discrete-time perturbed linear systems*

**Remark 14**. If the conditions of theorem 13 are verified, Θ is finitely determined, so Θ is described by a finite number of inequalities, and consequently Φ is also described by a finite number of inequalities.

In the next section, we will propose an algorithmic procedure to determine such integer $k_0$ described above.

## IV. Algorithmic Determination

Using the previous results, we can deduce an algorithm for computing the ϵ-capacity for a given gain matrix. This algorithm has almost the same theoretical convergence properties as the one proposed by Gilbert and Tan [7].

**Definition 15**. Let $k_0$ the smallest integer which satisfies $\Theta_{k_0} = \Theta$, $k_0$ is called the index of determination.

The last theorem and proposition 8 in the previous section, suggests the following algorithm for determining the index of determination $k_0$.

**Algorithm 1**:
___________________

*Step 1: Set $k = 0$*
*Step 2: if $\Theta_k = \Theta_{k+1}$*
    *$k_0 = k$ and stop*
*else continue*
*Step 3: $k = k + 1$ and return to Step2*
___________________

It is clear that the algorithm 1 will produce $k_0$ if and only if the ϵ-capacity Θ, of the chosen gain matrix K, is finitely determined. But, in the case where Θ is not finitely determined, there is no finite algorithmic procedure. Fortunately, as subsequent developments show, it is often possible to resolve the issue of finite determination by other means.

This algorithm can be considered as an initial version, because it does not describe how the test $\Theta_k = \Theta_{k+1}$ is implemented in Step 2. The main problem here is to achieve $\Theta_k = \Theta_{k+1}$. The difficulty can be overcome if we consider the following approach:
Let $\mathbb{R}^n$ be normed by

$$\|x\| = \max_{1 \leq i \leq n} |x_i|, \forall x = (x_1, \ldots, x_n) \in \mathbb{R}^n$$

Then, for every $k \geq 0$, the set $\Theta_k$ can be written down as follows:

$$\Theta_k = \{x \in \mathbb{R}^n / h_s(C\widetilde{A}^i x) \leq \epsilon, \forall i = 0, \ldots, k, \forall s = 0, \ldots, 2n\}$$

where $h_s: \mathbb{R}^n \to \mathbb{R}$ are such that for every $x \in \mathbb{R}^n$ and for every $j = 1, \ldots, n$ we have

$$\begin{cases} h_{2j-1}(x) = x_j - \epsilon \\ h_{2j}(x) = -x_j - \epsilon \end{cases}$$

**Remark 16**. We see that if Θ is finitely determined, then there exists the index of determination $k_0$ and Θ can be written as follows:

$$\Theta = \{x \in \mathbb{R}^n / h_s(C\widetilde{A}^i x) \leq \epsilon, \forall i = 0, \ldots, k_0, \forall s = 0, \ldots, 2n\}$$

From the remark 5, the test $\Theta_k = \Theta_{k+1}$ is equivalent to $\Theta_k \subset \Theta_{k+1}$ It follows that
$$\Theta_k = \Theta_{k+1} \Leftrightarrow \forall x \in \Theta_k, h_s(C\widetilde{A}^{k+1} x) \leq \epsilon, \forall s = 0, \ldots, 2n$$
or equivalently,

$$\sup_{x \in \Theta_k} h_s(C\widetilde{A}^{k+1} x) \leq \epsilon, \text{ for } s = 1, \ldots, 2n$$

Consequently the test $\Theta_k = \Theta_{k+1}$ leads to a set of mathematical programming problems. We will propose below another version of the algorithm 1, the modified algorithm is given as follows.

**Algorithm 2:**

*Step 1: Set $k = 0$*
*Step 2: Solve the following optimization problems for $s = 1, \ldots, 2n$*
*Maximize $J_s(x) = h_s(C\tilde{A}^{k+1}x)$*
*Subject to the constraints*
$$h_j(C\tilde{A}^l x) \leq \epsilon, \forall j = 1, \ldots, 2n, \forall l = 0, \ldots, k$$
*Let $J^*_s$ be the maximum value of $J_s(x)$*
*If $J^*_s \leq 0$ for $s = 1, \ldots, 2n$*
*Set $k_0 = k$ and define $\Theta$ by (10), and Stop*
*else continue*
*Step 3: Set $k = k + 1$ and return to Step2*

**Remark 17**.
*i)* This algorithm can never be useful, if there are no methods to solve rather large mathematical programming problems which arise in Step 2. This presents some difficulty because global optima are needed. Even when the functions $h_j, \forall j = 1, \ldots, 2n$ are convex, the difficulty remains because the programming problems require the maximum of a convex function subject to convex constraints.

*ii)* Assumptions of theorems 13 are sufficient but not necessary. If these conditions are not verified, there is no guarantee that Algorithm 2 will stop. If the Algorithm 2 converge then the $\epsilon$-capacity $\Theta$ is finitely determined, else it is not. In the case where $\Theta$ is not finitely determined, we can choose a different gain matrix K which ensures convergence of the algorithm and the assumptions of theorem 13. To illustrate this work we will give in the next section some numerical examples.

## V. Examples

Using the above ideas it is possible to characterize the $\epsilon$-capacity $\Theta$ of every given gain matrix K that verifies the necessary conditions for the convergence of the algorithm 2. In this section, we are content to apply them to some simple examples.

**Example1** : Consider the following system
$$\begin{cases} x_{i+1} = Ax_i + Bu_i \\ x_0 = \alpha \begin{pmatrix} 0{,}3 \\ 0{,}5 \end{pmatrix} + \beta \in \mathbb{R}^2 \end{cases}$$

where $(\alpha, \beta) \in \mathbb{R} \times \mathbb{R}^2$ is unknown and unavoidable disturbance which infect the initial state, $\tau_0 = \begin{pmatrix} 0{,}3 \\ 0{,}5 \end{pmatrix} \in \mathbb{R}^2$ and $A, B$ are described as follows :

$$A = \begin{pmatrix} 0{,}9 & 0 \\ 0{,}6 & 0{,}3 \end{pmatrix} \text{ and } B = \begin{pmatrix} -1{,}5 & 2 \\ 1 & -3 \end{pmatrix}$$

The associated output function is
$$y_i = Cx_i$$
where $C = (1\ 1)$ and $\epsilon = \sqrt{2}$. We can easily verify that the pair $(A, B)$ is controllable, so we can place the eigenvalues of $\tilde{A} = A + BK$ arbitrarily such that $\tilde{A}$ is asymptotically stable. For that, we choose K as follows
$$K = \begin{pmatrix} 0{,}32 & 0{,}16 \\ 0{,}24 & 0{,}12 \end{pmatrix}$$
which implies that
$$\tilde{A} = \begin{pmatrix} 0{,}9 & 0 \\ 0{,}2 & 0{,}1 \end{pmatrix}$$

Now, we are sure after a simple verification that the pair $(\tilde{A}, C)$ is observable. The theorem 13 ensure the convergence of the algorithm 2, which gives the index of determination $k_0 = 0$, then the $\epsilon$-capacity of the chosen matrix $K$ can given by

$$\Theta = \left\{ \begin{pmatrix} x \\ y \end{pmatrix} \in \mathbb{R}^2 / |x + y| \leq \sqrt{2} \text{ and } |1{,}1x + 0{,}1y| \leq \sqrt{2} \text{ and } |1{,}01x + 0{,}01y| \leq \sqrt{2} \right\}$$

and

$$\Phi = \left\{ K \in \mathbb{R}^{2\times 2} / \begin{pmatrix} 0,3 \\ 0,5 \end{pmatrix} \in \Theta \text{ and } \begin{pmatrix} 1 \\ 0 \end{pmatrix} \in \Theta \text{ and } \begin{pmatrix} 0 \\ 1 \end{pmatrix} \in \Theta \right\}$$

The figure 1 gives a representation of the set Θ corresponding to this example.

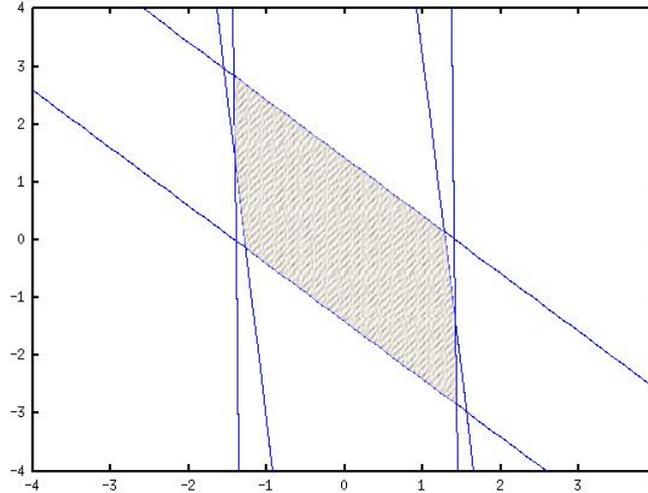

**Figure 1: The colored region represents the $\epsilon$-capacity Θ of the matrix $K = \begin{pmatrix} 0,32 & 0,16 \\ 0,24 & 0,12 \end{pmatrix}$ corresponding to example 1.**

We can by a simple hand calculation, see that $e_1 \in \Theta$ and $e_2 \in \Theta$ where $(e_i)_i, i = 1,2$ is the canonical basis of $\mathbb{R}^2$, which means that the disturbance β does not influence the associated output function. We can see also that $\begin{pmatrix} 0,3 \\ 0,5 \end{pmatrix} \in \Theta$ which means that the corresponding output function is not sensitive to the disturbance α. We can conclude that the selected gain matrix K have an admissible $\epsilon$-capacity, which means that $K$ belongs to Φ, which implies that the system is insensitive to all disturbances $(\alpha, \beta)$ under the associated control law $u_i = Kx_i$. Using the algorithm developed in the previous section, we give in Table 1 and in Table 2 the values of $k_0$ corresponding to different choices of the matrix defining the system.

| Ex | A | B | C | K | $\widetilde{A}$ | $\epsilon$ | $k_0$ |
|---|---|---|---|---|---|---|---|
| 2 | $\begin{pmatrix} 1 & 0 \\ 2 & 0,3 \end{pmatrix}$ | $\begin{pmatrix} -2 & 2 \\ 1 & -3 \end{pmatrix}$ | $(-1\ 1)$ | $\begin{pmatrix} 1,875 & 0,35 \\ 1,625 & 0,35 \end{pmatrix}$ | $\begin{pmatrix} 0,5 & 0 \\ -1 & -0,4 \end{pmatrix}$ | 0,4 | 1 |
| 3 | $\begin{pmatrix} 1 & 0 \\ -2 & 3 \end{pmatrix}$ | $\begin{pmatrix} -2 & 0 \\ -1 & 2 \end{pmatrix}$ | $(-0,1\ -1)$ | $\begin{pmatrix} 0 & 0 \\ 1,25 & -1,55 \end{pmatrix}$ | $\begin{pmatrix} 1 & 0 \\ 0,5 & -0,1 \end{pmatrix}$ | 0,1 | 3 |
| 4 | $\begin{pmatrix} 1 & -2 \\ 0,2 & 7 \end{pmatrix}$ | $\begin{pmatrix} -1 & 1 \\ 1 & 2 \end{pmatrix}$ | $(-0,9\ 1,9)$ | $\begin{pmatrix} 4,4 & 2 \\ -2,4 & -4 \end{pmatrix}$ | $\begin{pmatrix} -1 & 0 \\ -0,2 & 1 \end{pmatrix}$ | 0,2 | 1 |
| 5 | $\begin{pmatrix} 1 & 3 & 0 \\ -2 & 1 & 0 \\ 0 & 2 & 0 \end{pmatrix}$ | $\begin{pmatrix} -1 & -1 & 0 \\ 1 & 2 & -2 \\ 1 & 2 & 0 \end{pmatrix}$ | $(-0,1\ -0,3\ 0,2)$ | $\begin{pmatrix} 1 & 8,4 & 0 \\ -1 & -5,4 & 0 \\ -1 & -0,5 & 0 \end{pmatrix}$ | $\begin{pmatrix} 1 & 0 & 0 \\ -1 & -0,4 & 0 \\ -1 & -0,4 & 0 \end{pmatrix}$ | 0,2 | 4 |

**Table 1 : Data Of Examples 2-5**

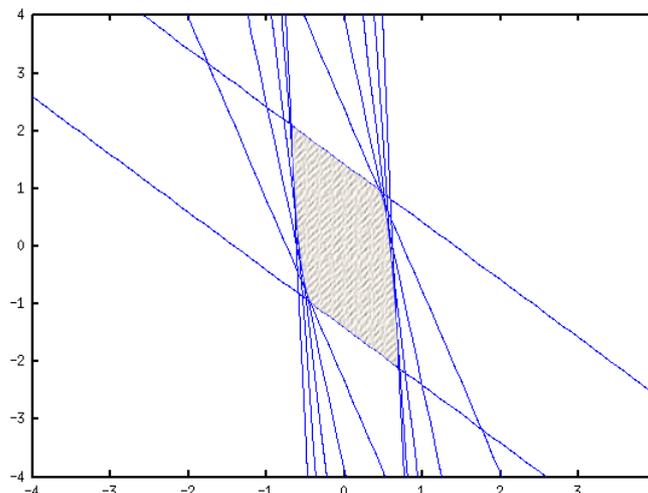

**Figure 2: The colored region represents the $\epsilon$-capacity Θ of the matrix K chosen in example 6.**

*The $\epsilon$-capacity of a gain matrix and tolerable disturbances: Discrete-time perturbed linear systems*

**Comment 1:** In example 1, the selected gain matrix K have an admissible $\epsilon$-capacity, the corresponding control law was able to reduce the effects of all disturbances $(\alpha, \beta) \in \mathbb{R} \times \mathbb{R}^2$, but for the same system, with another choice of K, example 6 shows that $e_2 \notin \Theta$, which means that the system is influenced by disturbances $\beta$, but not by $\alpha$ because $\begin{pmatrix}0,3\\0,5\end{pmatrix} \in \Theta$ (see Figure 2). In this case, the chosen K does not belong to $\Phi$, so it is not useful. In example 7, we can see also that the chosen control is not useful, because neither $\begin{pmatrix}0,3\\0,5\end{pmatrix} \in \Theta$ nor $e_2 \notin \Theta$ (see Figure 3).

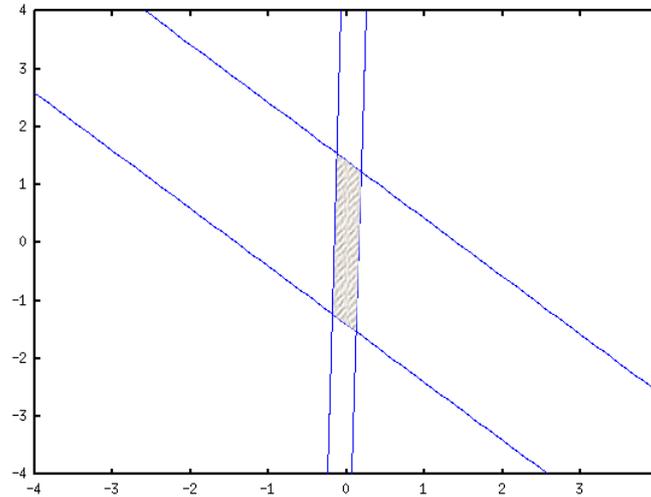

**Figure 3:** The colored region represents the $\epsilon$-capacity $\Theta$ of the matrix K chosen in example 7.

| Ex | A | B | C | $\widetilde{A}$ | $\epsilon$ | $k_0$ |
|---|---|---|---|---|---|---|
| 6 | $\begin{pmatrix}0,9 & 0\\0,6 & 0,3\end{pmatrix}$ | $\begin{pmatrix}-1,5 & 2\\1 & -3\end{pmatrix}$ | $(1\ 1)$ | $\begin{pmatrix}0,9 & 0\\0,99 & 0,6\end{pmatrix}$ | $\sqrt{2}$ | 5 |
| 7 | $\begin{pmatrix}0,9 & 0\\0,6 & 0,3\end{pmatrix}$ | $\begin{pmatrix}-1,5 & 2\\1 & -3\end{pmatrix}$ | $(1\ 1)$ | $\begin{pmatrix}1 & 0\\0,5 & -0,1\end{pmatrix}$ | $\sqrt{2}$ | 1 |
| 8 | $\begin{pmatrix}5 & 2\\4 & 0\end{pmatrix}$ | $\begin{pmatrix}7 & 1\\9 & 2\end{pmatrix}$ | $(-0,9\ 1,9)$ | $\begin{pmatrix}-1 & 0\\-0,2 & 1\end{pmatrix}$ | 0,2 | 1 |
| 9 | $\begin{pmatrix}1 & 1 & 0 & 3 & 5\\0 & 0 & 0 & 1 & 2\\1 & 5 & 3 & 5 & 3\\6 & 0 & 2 & 1 & 1\\0 & 3 & 5 & 0 & 3\end{pmatrix}$ | $\begin{pmatrix}1 & 1 & 0 & 3 & 5\\0 & 0 & 0 & 1 & 2\\1 & 5 & 3 & 5 & 3\\6 & 0 & 2 & 1 & 1\\0 & 3 & 5 & 0 & 3\end{pmatrix}$ | $\begin{pmatrix}-0,1 & -0,1 & 1 & 0 & -0,5\\-0,1 & -1 & 0 & 0 & 1\end{pmatrix}$ | $\begin{pmatrix}-1 & 0 & 0 & 0 & 0\\-1 & -1 & 0 & 0 & 0\\-1 & -0,4 & 0,3 & 0 & 0\\-1 & -0,4 & 0,3 & 1 & 0\\-1 & -0,4 & 0,3 & 0 & 2\end{pmatrix}$ | 0,9 | 32 |
| 10 | $\begin{pmatrix}0 & 4\\6 & 1\end{pmatrix}$ | $\begin{pmatrix}2 & 0\\0 & 1\end{pmatrix}$ | $\begin{pmatrix}2 & -2\\-1 & 0,04\\-1 & 3\end{pmatrix}$ | $\begin{pmatrix}1 & -1\\0 & -1\end{pmatrix}$ | 0,2 | 2 |

**Table 2 : Data Of Examples 6-10**

**Comment 2:** While the conditions in the theorem 13 are sufficient for the convergence of the algorithm 2, examples 3-5-7 show that they are not necessary. We can see that the matrix $\widetilde{A}$ is not asymptotically stable but only Lyapunov stable. Moreover, $\widetilde{A}$ is not stable in examples 4-8-9-10.

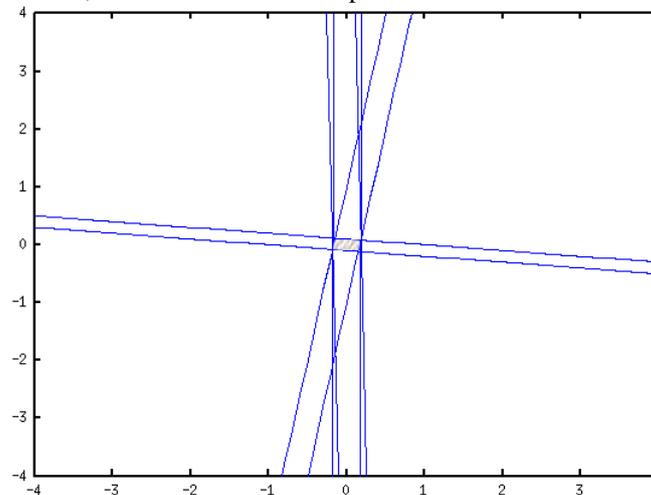

**Figure 4:** The colored region represents the $\epsilon$-capacity $\Theta$ of the matrix K chosen in example 3.

# VI. Conclusion

In this paper, we have studied the problem of discrete-time linear systems with perturbed initial state. The $\epsilon$-capacity of a gain matrix $K$ is introduced as a new approach to attenuate the effects of disturbances that infect the initial state, by the corresponding feedback control $u_i = Kx_i$. Necessary conditions have been obtained. Numerical examples were provided to show the effectiveness.

## References

[1]. Rakovic, S. V., Grieder, P., Kvasnica, M., Mayne, D. Q., & Morari, M.(2004, December). Computation of invariant sets for piecewise affine discrete time systems subject to bounded disturbances. In Decision and Control, 2004. CDC. 43rd IEEE Conference on (Vol. 2, pp. 1418-1423). IEEE.
[2]. Bouyaghroumni, J., El Jai, A., & Rachik, M. (2001). Admissible disturbance sets for discrete perturbed systems. APPLIED MATHEMATICS AND COMPUTER SCIENCE, 11(2), 349-368.
[3]. Cheng, Y., Xie, W., & Sun, W. (2012). High Gain Disturbance Observer-Based Control for Nonlinear Affine Systems. mechatronics, 1, 4.
[4]. Kunisch, K. (1988). A review of some recent results on the output least squares formulation of parameter estimation problems. Automatica, 24(4), 531-539.
[5]. Dahleh, M. A., & Shamma, J. S. (1992). Rejection of persistent bounded disturbances: Nonlinear controllers. Systems & control letters, 18(4), 245-252.
[6]. Wang, Z., Lam, J., & Liu, X. (2004). Exponential filtering for uncertain Markovian jump time-delay systems with nonlinear disturbances. Circuits and Systems II: Express Briefs, IEEE Transactions on, 51(5), 262-268.
[7]. Gilbert, E. G., & Tan, K. T. (1991). Linear systems with state and control constraints: The theory and application of maximal output admissible sets. Automatic Control, IEEE Transactions on, 36(9), 1008-1020.
[8]. Rachik, M., Abdelhak, A., & Karrakchou, J. (1997). Discrete systems with delays in state, control and observation: The maximal output sets with state and control constraints. Optimization, 42(2), 169-183.
[9]. Hirata, K., & Ohta, Y. (2008). Exact determinations of the maximal output admissible set for a class of nonlinear systems. Automatica, 44(2), 526-533.
[10]. Dórea, C. E., & Hennet, J. C. (1996). Computation of maximal admissible sets of constrained linear systems. In Proc. of 4th IEEE Med. Symposium (pp. 286-291).
[11]. Namerikawa, T., Shinozuka, W., & Fujita, M. (2004). Disturbance and Initial State Uncertainty Attenuation Control for Magnetic Bearings. In Proceedings 9th International Symposium on Magnetic Bearings (pp. 3-6).